\documentclass[aps,pre,groupedaddress]{revtex4}
\usepackage{graphicx,amsmath,amssymb,bm,multirow}

\newcommand{\eq}{\begin{equation}}
\newcommand{\feq}{\end{equation}}
\newcommand{\eqn}{\begin{eqnarray}}
\newcommand{\feqn}{\end{eqnarray}}
\newcommand{\arr}{\begin{eqnarray*}}
\newcommand{\farr}{\end{eqnarray*}}
\newcommand{\beq}{\begin{equation}}
\newcommand{\eeq}{\end{equation}}
\newcommand{\bea}{\begin{eqnarray}}
\newcommand{\eea}{\end{eqnarray}}
\newcommand{\lb}{\label}

\def\Y{Yakushevich}

\def\a{\alpha}

\def\bho{d}

\def\kms{{\hbox{km/s}}}
\def\Nm{{\hbox{N/m}}}
\def\kJm{{\hbox{kJ/mol}}}
\def\eV{{\hbox{eV}}}

\def\varphin{\varphi_{n,i}}

\numberwithin{equation}{section}

\begin{document}

\title{Numerical analysis of solitons profiles\\ in a composite model for DNA torsion dynamics}

\author{Roberto De Leo}
\email{roberto.deleo@ca.infn.it} \affiliation {Dipartimento di
Fisica, Universit\`a di Cagliari \\ and INFN, Sezione di Cagliari
\\ Cittadella Universitaria, 09042 Monserrato, Italy}
\author{Sergio Demelio}
\email{sergio.demelio@ca.infn.it} \affiliation {Dipartimento di
Fisica, Universit\`a di Cagliari \\ and INFN, Sezione di Cagliari
\\  Cittadella Universitaria, 09042 Monserrato, Italy}

\begin{abstract}\noindent
We present the results of our numerical analysis of a ``composite''
model of DNA which generalizes a well-known elementary torsional model of 
Yakushevich by allowing bases to move independently from the backbone.
The model shares with the \Y\ model many features and results but 
it represents an improvement from both the conceptual and the phenomenological
point of view. It provides a more realistic description of DNA and
possibly a justification for the use of models which consider the
DNA chain as uniform. It shows that the existence of solitons is a
generic feature of the underlying nonlinear dynamics and is to a
large extent independent of the detailed modelling of DNA. 
As opposite to the \Y\ model, where it is needed to use an unphysical
value for the torsion in order to induce the correct velocity of sound,
the model we consider supports solitonic solutions, qualitatively and
quantitatively very similar to the \Y\ solitons, in a fully
realistic range of all the physical parameters characterizing the
DNA.
\end{abstract}
\maketitle
\section{Introduction}
It is now almost thirty years that the role of solitons in fundamental
DNA functions is under serious investigation (see~\cite{CDG07,CDDG07} for 
a long list of references).
In particular a lot of efforts have been made lately~\cite{R93,HmK04a,HmK04b} 
to study the phenomenon of the so-called {\sl base-flipping}, namely the 
complete opening of a narrow segment of DNA, a phenomenon which is thought 
to be important for fundamental processes such as replication and
transcription. 
\par
A successful elementary model for DNA rotational modes, introduced by 
Yakushevich~\cite{Yak04}, makes use of a double chain of oscillators 
where every node of each chain is represented by a disc which
interacts with their first neighbors on the same chain and with the disc 
in front of it on the other chain.
%
\begin{figure}
  \includegraphics[width=200pt]{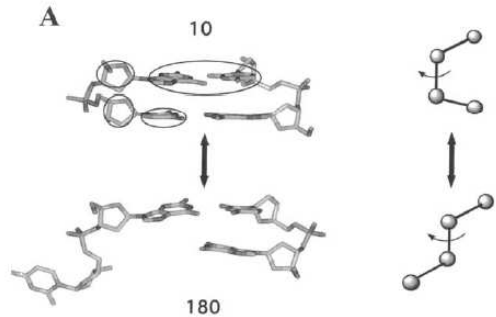}
  \\
  \caption{A picture (extracted from~\cite{HmK02}) showing the phenomenon of {\sl base-flipping},
  namely the complete opening of a DNA's base pair due to the rotation of (at least) on of them 
  by an angle of $\pi$. It is clear from the picture that in the process of rotating about the
  DNA axis both the sugar and base planes also rotate about some other axis; to simplify as much
  as possible our model we will disregard this effect and concetrate only on the rotation about
  DNA's axis.}
  \label{fig:bf}
\end{figure}
The homogeneous version of this model succeds in supporting the existence 
of topological solitons for a wide range of the physical and geometrical 
parameters, also allowing analytical solutions for a few cases, but it
was shown~\cite{Yak04} that adding enough inhomogeneities -- i.e. taking
into account the geometrical and dynamical differences between the
four possible bases for a realistic DNA segment -- causes solitons
to lose energy and finally stop their motion after a few nodes on the chain.
\par
Our final goal, which will not be reached within this paper, is to determine
whether or not realistic segments of DNA support the motion of ``rotational'' 
solitons. In the composite model proposed in~\cite{CDG07}, by separating the degrees 
of freedom of the bases from the backbone-sugar component, we produced a model where 
the component supporting the solitons, i.e. the backbone-sugar chain, is perfectly 
homogeneous and the bases inhomogeneities act just as a perturbation of a homogenous 
system, leaving hope for a longer life of solitons moving on it.
We leave to a future paper the study of the profiles of solitons in inhomogeneous chains
and their time evolution.
\par
The numerical results we present in this paper support our guess that
the composite model represents an important improvement of the 
simple torsional models for DNA dynamics in the homogeneous approximation:
it supports the existence of solitons 
and moreover these solitons are very close to the corresponding ones 
in the \Y\ model. As a bonus, the increase in the geometrical (and consequentely
dynamical) detail turned out to be enough to allow the generation 
of the solitons corresponding to the \Y\ ones using for the
coupling constants values which are compatible with the physical 
ones~\footnote{In~\cite{Yak04} \Y\ needs to set the torsional energy to the unphysical
value of $6000 \kJm$ in order to induce the correct speed of sound in DNA $v\simeq1$ \kms}.
\par
As a final remark, we point out that all numerical results and estimates of the 
geometrical and dynamical parameters included in this paper are an improvement
and/or an update of the corrisponding ones published in~\cite{CDG07}, which is fully
superseded by this one.
\section{Model \& equations}
\label{sec:model}
\begin{figure}
  \begin{tabular}{cc}
    \includegraphics[width=200pt,bb = 170 450 450 710]{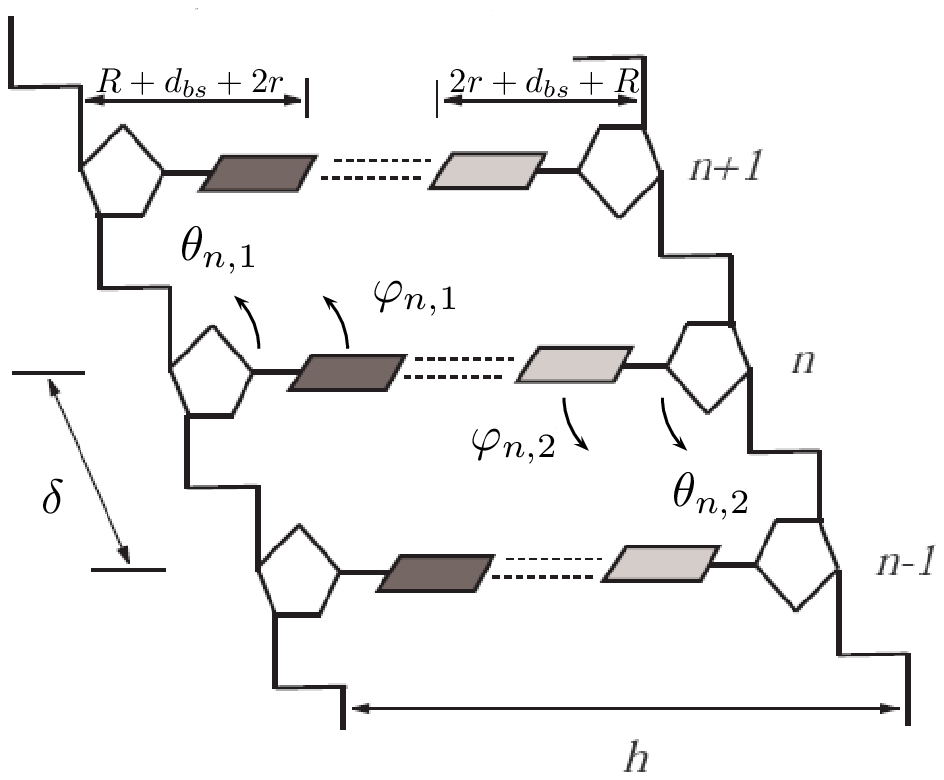}\ 
    \includegraphics[width=300pt,bb = 110 610 502 762]{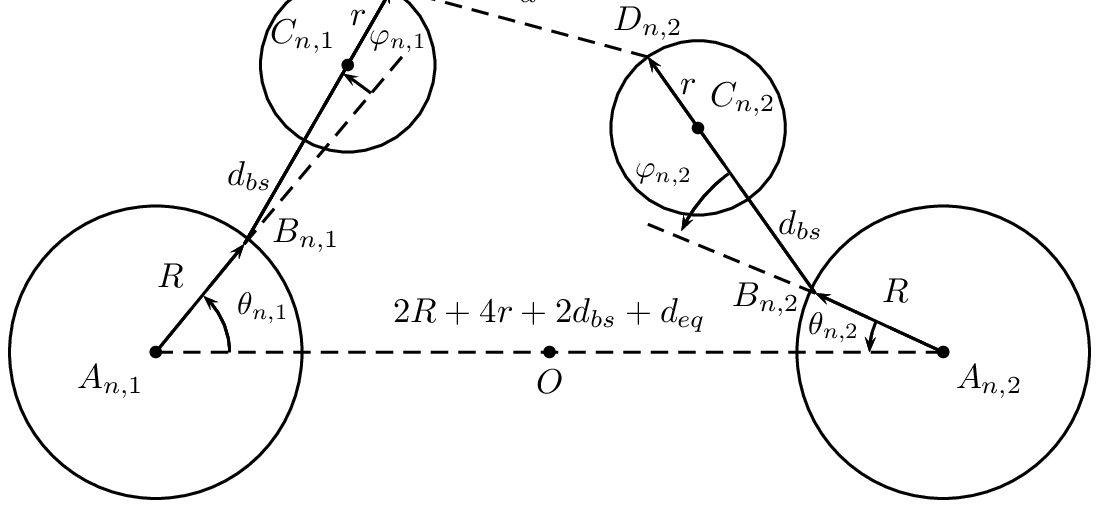}
  \end{tabular}
  \\
  \caption{
    {\sl Left.} A fragment of our ``composite'' model of the DNA's double chain: 
    at every node of the chain there are four degrees of freedom, two for each base 
    (the $\{\varphi_{n,i}\}$) and two for each sugar-phosphate group 
    (the $\{\theta_{n,i}\}$). 
    {\sl Right.} Detail of a chain node. Every base $\{n,i\}$ is allowed to rotate 
    about the $C_1$ atom of the corresponding sugar (see fig.~\ref{fig:atgc}), represented 
    by the point $B_{n,i}$, only by an angle between $[-\pi/2,\pi/2]$ because of the physical 
    contraint represented by the sugar pentagon, so that the effective phase space of the 
    model is $({\mathbb S}^1\times[-\pi/2,\pi/2])^{2n}$; this steric constraint is implemented 
    in the model dynamically through an effective potential. See sec.~\ref{sec:numerology} 
    for an evaluation of all geometrical constants that appear in the pictures above.
    }
  \label{fig:dna}
\end{figure}
%
Our model for DNA is a homogeneous double chain of coupled double pendula which
is a natural generalization of a well known model by \Y\  where, at every node of each chain, 
the whole group base-sugar-phosphate is represented by a single disc centered at the chain's backbone
axis: we simply split the group in two distinct discs, one still centered about the backbone 
axis and representing the sugar-phosphate group and the second rigidly rotating about a fixed
point of the former as shown in fig.~\ref{fig:dna} (note that, in order to keep the geometry of
the model as simple as possible, we consider the chain as plane
rather than helicoidal).
\par
Referring to fig.~\ref{fig:dna}, the coordinates of the two discs centers and of the extrema
of bases, whose distance we use to determine the intensity of the bonds between the base 
pairs on the same chain node, are the following:
$$
\vbox{
  \halign{$\displaystyle #$\hfill&$\displaystyle #$\hfill&$\displaystyle #$\hfill&$\displaystyle #$\hfill\cr
    A_{n,i}&\big((-1)^ih/2&,0&,n\delta\big)\cr
    C_{n,i}&\big((-1)^ih/2+(-1)^{i+1}\left(R\cos\theta_{n,i}+(d_{bs}+r)\cos(\theta_{n,i}+\varphi_{n,i})\right)&
    ,(-1)^{i+1}\left(R\cos\theta_{n,i}+(d_{bs}+r)\cos(\theta_{n,i}+\varphi_{n,i})\right)&,n\delta\big)\cr
    D_{n,i}&\big((-1)^ih/2+(-1)^{i+1}\left(R\cos\theta_{n,i}+(d_{bs}+2r)\cos(\theta_{n,i}+\varphi_{n,i})\right)&,
    (-1)^{i+1}\left(R\cos\theta_{n,i}+(d_{bs}+2r)\cos(\theta_{n,i}+\varphi_{n,i})\right)&,n\delta\big)\cr
  }
}
$$
where $h=2R+4r+2d_{bs}+\bho_{eq}$.
\par
The dynamical evolution of our mechanical system is  determined by the Lagrangian $L=T-V$, 
where $T$ is the kinetic energy and $V$ is the interaction potential.
The interactions that are relevant for DNA rotational dynamics are five:
\begin{enumerate}
\item{} The {\sl torsion} ($V_t$) between next neighbor sugar-phosphate groups on the same chain, 
representing the torsional elasticity of the backbone. This force is the results of complex molecular
interactions at the backbone level and it is to be considered an ``effective'' term. Following the principle
of keeping the potential expressions as simple as possible until some good reason is found to make them
more complicated and keeping in mind that $\theta$ can vary on the whole circle, we use for it the 
``physical pendulum'' periodic potential 
\eq
V_t = \sum_{n=1}^{N-1}\sum_{i=1}^2 K_t \left[1-\cos(\Delta\theta_{n,i})\right]
\feq
where $\Delta\theta_{n,i}=\theta_{n+1,i}-\theta_{n,i}$.
Note that relative angles between next neighbor discs never get big, so that it
would be safe also to use its harmonic approximation $Vt\simeq \sum_{n=1}^{N-1}\sum_{i=1}^2K_t/2[\Delta\theta_{n,i}]^2$.

\item{} The {\sl stacking} ($V_s$) between next neighbors bases on the same chain, 
representing the $\pi - \pi$ bonds between the rings that constitute the bases.
This interaction is much better understood than the previous one and in particular
it is clear that only depends on the relative displacement between next neighbor bases,
going rapidly to zero together with the overlapping portion of their surface, 
e.g. like in a Morse-like potential. However, since also in this
case the relative angle of any two bases next to each other keeps small,
we simplified its expression by considering it a harmonic bond depending on the ``$xy$''
distance between the centers of the bases:
\eqn
\label{eq:Vs}
 &V_s&= \displaystyle\sum_{n=1}^{N-1}\sum_{i=1}^2 \frac{1}{2}K_s\;d^2_{xy}(C_{n+1,i},C_{n,i})\cr
 &&= \displaystyle\sum_{n=1}^{N-1}\sum_{i=1}^2 \frac{1}{2}K_s(d_{bs}+r)^2
  \bigg\{
    \left[
          \alpha(\cos\theta_{n+1,i}-\cos\theta_{n,i})+\cos(\theta_{n+1,i}+\varphi_{n+1,i})-\cos(\theta_{n,i}+\varphi_{n,i})
    \right]^2
      +
      \cr
      &&\phantom{= \displaystyle\sum_{n=1}^{N-1}\sum_{i=1}^2 \frac{1}{2}K_s(d_{bs}+r)^2\bigg\{[}
    \left[
          \alpha(\sin\theta_{n+1,i}-\sin\theta_{n,i})+\sin(\theta_{n+1,i}+\varphi_{n+1,i})-\sin(\theta_{n,i}+\varphi_{n,i})
    \right]^2
  \bigg\}
\feqn
where $\alpha=R/(d_{bs}+r)$.
\item{} The {\sl pairing} ($V_p$) between bases on opposite chains, representing the
ionic bonds which keep the helices together. This is the best understood force among
the ones we are considering and, like in the stacking case, it is known to go rapidly 
to zero a few Angstrom far from the equilibrium position; in this case though the distance 
between pairs of bases does get big when a base is flipping and therefore a harmonic 
approximation result rather unphysical. The interaction can  be more realistically 
modelled by a Morse-like potential~\cite{GaeNMP}, nevertheless we will produce profiles also in 
the harmonic approximation in order to compare our results with those in~\cite{Yak04},
so we will consider both cases:
\eqn
\label{eq:Vp}
  &V_p &= \displaystyle\sum_{n=1}^{N} D\left[ 1-e^{-\mu\;d(D_{n,1},D_{n,2}}) \right]^2\cr
  &&\simeq \displaystyle\sum_{n=1}^{N}\frac{1}{2} K_p(d_{bs}+2r)^2
  \bigg\{
    \left[
          \beta(\cos\theta_{n,1}+\cos\theta_{n,2})+\cos(\theta_{n,1}+\varphi_{n,1})+\cos(\theta_{n,2}+\varphi_{n,2})-2-2\beta
    \right]^2
      +\cr
      &&\phantom{= \displaystyle\sum_{n=1}^{N}\frac{1}{2} K_p(d_{bs}+2r)^2\bigg\{[}
    \left[
          \beta(\sin\theta_{n,1}+\sin\theta_{n,2})+\sin(\theta_{n,1}+\varphi_{n,1})+\sin(\theta_{n,2}+\varphi_{n,2})
    \right]^2
  \bigg\}
\feqn
where $\beta=R/(d_{bs}+2r)$ and $K_p=2 D \mu^2$ (see eq.~(\ref{morse})). Note that, in order to simplify the expression of the elongation from the
equilibrium position, we made the ``contact'' approximation $\bho_{eq}=0$, i.e. we disregarded the inter-bases 
distance in the equilibrium position; we will show numerically in sec.~\ref{sec:numerics} that this does not change 
significantly the solitons profiles. The same is knonw to hold also in the \Y\ model.
\item{} The {\sl helicoidal} interactions ($V_h$) between nucleotides, which are mediated by water filaments 
(Bernal-Fowler filaments). This is the only ingredient of our model which is reminiscent of the helicoidal 
structure of DNA; in particular we will consider those being on opposite helices at half-pitch distance, as they are near
enough in three-dimensional space due to the double helical geometry. As the nucleotide move, the hydrogen bonds
in these filaments – and those connecting the filaments to the nucleotides – are stretched and thus resist differential
motions of the two connected nucleotides.
We will, for the sake of simplicity and also in view of the small energies involved, only consider filaments forming
between the sugar-phosphate groups, thus only the backbone angles will be involved in these interactions. 
Recalling that the pitch of the helix corresponds to 10 bases in the B-DNA equilibrium configuration we set
\eq
V_h = \sum_{n=1}^{N}\sum_{i=1}^2 K_h \left[ 1-\cos(\theta_{n+5,i+1}-\theta_{n,i}) \right]
\feq
where the sum $i+1$ is meant {\sl modulo} 2.
\item{} The {\sl sugar wall} ($V_{sw}$), representing an ``effective'' interaction which dynamically restricts 
the range of the base angles $\varphi_{n,i}$ to some interval $[\varphi_-,\varphi_+]$ -- ruling out this way 
the possibility of topologically non trivial configurations for them -- to represent the steric constraint 
represented by the sugars, which prevents the corresponding bases from doing a complete circle about them. 
We set these angles to $\phi_\pm=\pm\pi/2$ and verified numerically that the profiles in the composite
model converge to the corresponding ones by narrowing more and more the interval $[\varphi_-,\varphi_+]$.
In order not to interfere with the dynamics close to the equilibria positions the potential must be as flat
as possible close to the zeroes of the $\varphi_{n,i}$ and must rise rather quickly when the $\varphi_{n,i}$
approach $\pm\pi/2$. A natural choice, which we implemented in~\cite{CDG07}, would be to use some high power 
of the {\sl tangent} function but divergences cause problem in numerical calculations so 
we  use in this paper  some high even power of the {\sl sine} function.
\eq
\label{eq:sw}
V_{sw} = \sum_{n=1}^{N}\sum_{i=1}^2 K_{sw} \sin^k(\varphi_{n,i})
\feq
where the coupling constant $K_{sw}$ must be taken big enough to prevent the bases from passing through the 
$\pm\pi/2$ barrier but also not so big to interfere too much with the dynamics when the $\varphi_{n,i}$
are closer to the equilibrium position.
\end{enumerate}
As for the kynetic energy, a straightforward calculation shows that for the sugar-phosphate group we have
\eq
T_t = \sum_{n=1}^N\sum_{i=1}^2 \frac{1}{2}I_t \dot\theta^2_{n,i}
\feq
and for the bases
\eq
T_s = \sum_{n=1}^N\sum_{i=1}^2 \frac{1}{2}I_s 
  \left[ \dot\varphi^2_{n,i} + 2(1+\alpha\cos\varphin)\dot\varphi_{n,i}\dot\theta_{n,i} + (1+2\alpha\cos\varphin+\alpha^2)\dot\theta^2_{n,i} \right]
\feq
Note that by putting the bases angles identically equal to zero the Lagrangian $L=T_t+T_s-V_t-V_s-V_p-V_h-V_{sw}$
of the composite model reduces, modulo the helicoidal term, exactly to the \Y\ homogeneous Lagrangian:
\eqn
\label{eq:gyak}
L(\theta_{n,i},0,\dot\theta_{n,i},0)=&\phantom{+}\displaystyle\sum_{n=1}^N\sum_{i=1}^2 \frac{1}{2}\left[I_t+(1+\alpha)^2I_s\right] \dot\theta^2_{n,i}
+\displaystyle\sum_{n=1}^{N-1}\sum_{i=1}^2\left[K_t+(1+\alpha)^2(d_{bs}+r)^2K_s\right] \left[1-\cos(\Delta\theta_{n,i})\right]\cr
&+\displaystyle\sum_{n=1}^N (1+\beta)^2(d_{bs}+2r)^2K_p\left\{2(1-\cos\theta_{n,1})+2(1-\cos\theta_{n,2})-\left[1-\cos(\theta_{n,1}-\theta_{n,2})\right]\right\}\cr
\feqn
\par
Now, once an initial condition for all angles is given, the Lagrangian determines completely the evolution 
of the state as the ``trajectory'' $q(t)=(\theta_{n,i}(t),\varphi_{n,i}(t))$ extremizing
the action $I=\int_{q(t)}L(q(t),\dot q(t))dt$. The initial states we are interested in are those ones which 
give rise to motions which do not change (much) shape, i.e. which move ``rigidly'', satisfying therefore the 
``discrete wave'' condition $q_{n+k,i}(t)=q_{n,i}(t-k\delta/v)$, where $\delta$ is the distance between 
two consecutive nodes. 
We look at these solutions as ``discrete solitons'' able to move on the DNA chain.
\par
For a discussion about the analytical properties of the model we refer the reader to the paper~\cite{CDDG07}. 
Even in the simplest cases it is impossible to find an explicit solution 
to the Lagrange equations and therefore the system must be analyzed numerically.
To this end we use the fact that in our case $h=-\delta/v$ is small enough to claim that
$$
\dot q_{n,i}(t) \simeq \frac{q_n(t-h)-q_n(t)}{-h} = -\frac{v}{\delta}[q_n(t+\delta/v)-q_n(t)] = -\frac{v}{\delta}\Delta_{n,i}[q(t)]
$$
so that the kynetic energy becomes
$$
T_t = \sum_{n=1}^N\sum_{i=1}^2 \frac{I_t v^2}{2\delta^2} [\Delta\theta_{n,i}]^2
$$
$$
T_s = \sum_{n=1}^N\sum_{i=1}^2 \frac{I_sv^2}{2\delta^2} 
\left\{ [\Delta\varphi_{n,i}]^2 + 2(1+\alpha\cos\varphin)\Delta\varphi_{n,i}\Delta\theta_{n,i} + (1+2\alpha\cos\varphin+\alpha^2)[\Delta\theta_{n,i}]^2 \right\}
$$
Having discretized time we are left with just one degree of freedom, the discrete space coordinate $n$,
so that what was previously the system lagrangian $L = \sum_{n=1}^N\sum_{i=1}^2 L_{n,i}$ has become now 
the action (summed over $n$) of the discrete Lagrangian $L_n=\sum_{i=1}^2L_{n,i}$; the discrete Lagrange equations, by the minimum
action principle, are clearly given simply by $dL=(\partial_{\theta_{n,i}}L,\partial_{\varphi_{n,i}}L,)=0$ and its solutions,
subjected to opportune initial conditions,
are exactly the profiles of the solitons that we want to make sure to be present in this generalization of the \Y\  model.
\section{Physical values of parameters \& dispersion relations}
\label{sec:numerology}
The evaluation of the main geometrical and dynamical physical quantities 
entering in our model is far from being a trivial matter.
There are still very few direct measurements, for example, of the stacking energy between bases within DNA
and the torsional backbone force -- due to quite complex interactions between the sugar and the phosphate
atoms -- is just an {\sl effective} term for which it is hard in principle to define a strenght.
Also for geometrical parameters things are not better and often in the literature different values for the same physical 
quantities are used. In this section we are going to give an estimate for the values of the parameters of our model.
%
\subsection{Kinematical parameters}
The kinematical parameters include the geometrical parameters, the mass $m$ and the momenta of inertia $I$.
We opted to directly estimate them starting
from PDB data~\cite{PDBRep} (note, as already mentioned, that the present results 
supersede the ones we used in~\cite{CDG07})
%
\begin{figure}
  \includegraphics[width=450pt]{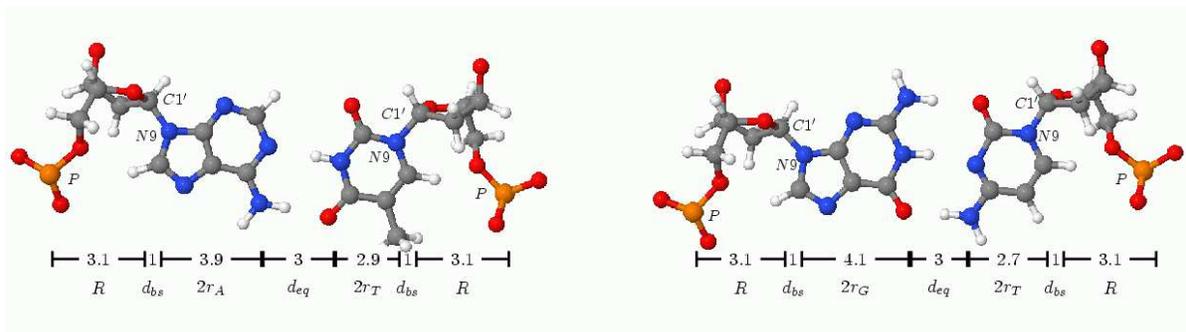}
  \caption{%
    A node of the DNA chain with a AT pair (left) and a GC pair (right). The measures
    of the distances are meant as the projections of the actual lenghts over the
    axis connecting the two phosphates of the same chain node. The distance between
    the two phosphates is $h=18$\AA\ for both base pairs.
  }
  \label{fig:atgc}
\end{figure}
%
The masses can be easily calculated on the basis of the chemical structure of DNA.
Estimate of momenta of inertia is more involved, especially for the sugar-phosphate group,
since the evaluation depends on the choice of the rotation axis and, as pointed out earlier,
base-flipping rotations are a complex subject.
In our evaluations we made the simplest approximation, namely momenta of inertia of sugar 
rings have been calculated with respect to rotations about the main axis of DNA
and passing by $P$ atom of the backbone (see fig.~\ref{fig:atgc})
and inertia momenta of bases have been calculated with respect to rotations about the same axis 
by $C_1$ atom of the sugar ring.
\par
The geometrical parameters of interest for our model are the longitudinal width of bases
$l_{b}$ and of the sugar $l_{s}$, the  distances of the bases from the relative sugars
$d_{bs}$ and the distance of a base from the relative dual base $d_{eq}$. 
We give our estimates for the masses, moments of inertia and the geometrical parameters 
for the different bases and their mean values in table~\ref{tab:kynParms}. From those data and
using the notations $R = l_{s}$, $r = {\bar l}_d/2$, $h = l_{s} + d_{bs} + {\bar l}_{b} + d_{eq}/2$, 
$h_c = l_{s} + d_{bs} + {\bar l}_{b}$ (the bar denotes the mean value), we get the average 
values for the geometrical parameters appearing in our Lagrangian shown in table~\ref{tab:para}.
\begin{table}[ht]
\centering
 \begin{tabular}{|l|c|c|c|c|c|c|}
  \hline
    & A & T & G & C & mean & Sugar\\
    \hline
  $m$ & 134  & 125  & 150  & 110  & 130 & 85 \\
  $I$ & $3.6\times10^3$ & $3.0\times10^3$ & $4.4\times10^3$ & $2.3\times10^3$ & $3.3\times10^3$ & $2.9\times10^3$\\
  $l$ & $3.9$ & $2.9$ & $4.1$ & $2.7$ & $3.4$ & $3.1$\\
  $d_{bs}$ & $1.0$ & $1.0$ & $1.0$ & $1.0$ & $1.0$ & -\\
  $d_{eq}$ & $3.0$ & $3.0$ & $3.0$ & $3.0$ & $3.0$ & -\\
 \hline
\end{tabular}
\caption{
  Order of magnitude for the basic geometrical parameters of the DNA.
  Units of measure are: atomic unit for masses $m$, $1.67 \times 10^{-47}
  {\rm Kg}\cdot {\rm m}^2$ for the inertia momenta $I$, Angstrom for $l$ (the
  longitudinal width of bases and sugar), $d_{bs}$  (the distances sugar-base) 
  and $d_{eq}$ (the distance at the equilibrium for the pairs AT and GC). 
  These values have been extracted from the sample ``generic'' B-DNA PDB data~\cite{PDBRep},
  provided by the Glactone Project~\cite{Glac}, and double checked with the data from
  \cite{BDNA81}, that agrees within $5\%$. As shown in fig.~\ref{fig:atgc}, the lengths
  of bases and sugar were taken by projecting them on the direction passing
  through the two phosphate atoms of the chain node to  best fit the geometry
  of the model; the (real) measure for the diameters of the bases A, T, G, C, 
  and the sugar and for $d_{bs}$ are respectively (in \AA): 
  4.6, 4.0, 5.7, 4.0, 3.3 and 1.5.) 
}
 \label{tab:kynParms}
\end{table}
\begin{table}[ht]
\centering
   \begin{tabular}{|c|c|c|c|c|c|c|c|}
    \hline
    $R$ & $r$ & $d_{bs}$ & $\bho_{eq}$ & $h$ & $h_c$ & $\alpha$ & $\beta$\\
    \hline
    3.1  &1.7  & 1.0 & 3.0 & 18 & 15 & 1.1 & 0.7  \\
    \hline
   \end{tabular}
   \caption{%
     Numerical values (in \AA) of the geometrical parameters characterizing our model.
     Here $h=2R+4r+2d_{bs}+\bho_{eq}$ is the width of DNA and $h_c=h|_{\bho_{0}=0}$ is
     its value in the contact approximation.
   }
   \label{tab:para}
\end{table}
%
%
\subsection{Dynamical parameters}
The determination of the four coupling constants ($K_p$, $K_t$, $K_s$ and $K_h$) 
characterizing our model requires some assessments. Their order of magnitude can be 
estimated by considering the typical energies of hydrogen bonds ($K_p$) and the 
experimental results for the torsional rigidity of the chain ($K_t$ and $K_s$). 
As we will show later, the geometry of our composite model allows
-- with values for coupling constants {\it within} the experimental ranges --
to make predictions about dynamical properties of DNA,
as the induced optical frequencies and the phonon velocities,
of the same order of magnitude of those experimetally observed
(conversely, dispersion relations allow 
to restrict the selection of the coupling constants
to those values which fit better our model).
\subsubsection{Pairing}
The coupling constant $K_p$ of the pairing potential (\ref{eq:Vp}) can be determined 
by considering the typical energy of hydrogen bonds. The pairing interaction involves
two (in the $A-T$ case) or three (in the $G-C$ case) ionic hydrogen bonds. 
We can model the pairing potential with a Morse function like
\beq \lb{morse} V_p(d)=D[1-e^{-\mu(\bho-\bho_{eq})}]^2
=\frac{1}{2}K_p(\bho-\bho_{eq})^2+O(\bho^3)\,,
\feq 
where $D$ is the potential depth, $\bho$ the distance between the points $D_{n,1}$ and 
$D_{n,2}$ (see fig.~\ref{fig:dna}), $\bho_{eq}$ the bond equilibrium distance
and $\mu$ a parameter that controls the width of the well. 
%
\begin{figure}
  \includegraphics[width=150pt]{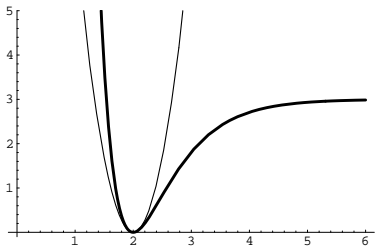}
  \\
  \caption{Comparison of Morse potential and its correspondig
    harmonic approximation}
  \label{fig:ma}
\end{figure}
Different estimates  of the parameters appearing in the potential
(\ref{morse}) are present in the literature. The estimates
$$ D_{AT} = 0.030\, \eV , \ D_{GC} = 0.045\, \eV, \
\mu_{AT} = 1.9\, {\rm \AA}^{-1}, \ \mu_{GC} = 2.5\, {\rm \AA}^{-1}
$$ are given in \cite{CP92} and used in \cite{ZC95}.
The values
$$ D = 0.040\, \eV, \ \mu = 4.45\, {\rm \AA}^{-1} $$ are given in
\cite{PBD92} and used in \cite{PBD92,BCP99,DeLuca04}. Finally,  the
estimates
$$ D_{AT} = 0.050\, \eV, \ D_{GC} = 0.075\, \eV, \
\mu_{AT} = \mu_{GC} = 4\, {\rm \AA}^{-1} $$ are given in \cite{Campa00}
and used in \cite{Campa00,KS05}. The values of coupling constants
corresponding to these different values for the parameters
appearing in the Morse potential range across an order of
magnitude: \beq\lb{kd} 3.5\, \Nm \ \leq \ K_p \, := \, 2
D \mu^2 \ \leq \ 38\, \Nm \ . \feq
In our numerical investigations we will use a value of $K_{p}$
near to the lower bound given in (\ref{kd}); that is,
we adopt the value $K_p = 4\, \Nm$ 
-- corresponding to $ D = 0.030\, \eV, \ \mu = 2\, {\rm \AA}^{-1}$ --
which leads to an optical excitation threshold of 
$q_4=\sqrt{2K_p/m_b}\simeq32\, {\rm cm}^{-1}$ (see eq.~(\ref{speed1})), 
so to be in agreement with~\cite{Pow87}.
\subsubsection{Stacking}
The determination of the torsion and stacking coupling constants
is more challenging and is based on a smaller amount of experimental
data. The main information is the total torsional rigidity of the
DNA chain  $C=S \delta$, where $\delta=3.4 {\rm \AA}$
is the base-pair spacing and $S$ is the torsional rigidity. 
It is known~\cite{BZ79,BFLG99} that 
\beq
10^{-28}\, {\rm J\cdot m} \ \leq \ C \ \leq 4 \cdot 10^{-28} \, {\rm J}\cdot{\rm m} \ . 
\feq 
This information is used e.g. in \cite{Eng80,Zhang89}, whose estimate 
is based on the evaluation of the free energy of superhelical winding; 
this fixes the range for the total torsional energy to be 
\beq 
180 \, \kJm \ \leq \ S \ \leq \ 720 \, \kJm \ . 
\feq
%
%
In our composite model the total torsional energy of the DNA chain
has to be considered as the sum of two parts, the base stacking
energy and the torsional energy of the sugar-phosphate backbone.
In order to extract the stacking coupling constant we use the
further information that $\pi-\pi$ stacking bonds amount at the
most to $50 \kJm$ \cite{HS90, Khair}. 
The stacking potential can be described by a Morse potential
whose width $\sigma$ is of the order 1\AA\ 
-- so that (see eq.~(\ref{eq:Vs})) the harmonic approximation is justified -- 
and assuming for the stacking energy the highest value possible, 
the coupling constant value is $K_s=2E_s/\sigma^2\simeq 16.6 \Nm$.
The phonon speed induced by this value of $K_s$ is 
$c_3 = \delta \sqrt{K_s/m_b} \simeq 3 \kms$ (see eq.~(\ref{speed1})),
which is rather close to the the estimate of 
$1.8 \kms \leq c_1 \leq 3.5 \kms$ given in \cite{Yak04}.
%
%
\subsubsection{Torsion and helicoidal couplings}
\label{sec:hel}
Once picked up the stacking component, our estimate for the
torsional coupling constant $K_t$ falls in the range 
\beq 
130 \, \kJm \ \leq \ K_t \ \leq \ 670 \, \kJm \ . 
\feq
Assuming (see~\cite{GaeJBP}) that $K_h \simeq K_t/25 = 5\, \kJm$, 
so that $c_4=\sqrt{2K_t/I_s}$ (see equations (\ref{speed1}) and (\ref{speed})),
all of these values for $K_t$ induce phonon speeds slightly higher
with respect to the estimates cited above, between 5 km/s and 11
km/s. For our numerical investigations, to keep the phonon speed
as low as possible, we will  set $K_t=130\, \kJm$.
\subsubsection{Sugar wall}
\label{sec:sw}
In order to have a barrier flat enough close to the equilibrium position 
we chose to use the exponent $k=100$ in~\ref{eq:sw}. Numerical tests 
show that a value of at least $K_{sw}=10^4\, \kJm$ is needed in order
to keep the bases angles within the $[-\pi/2,\pi/2]$ interval.
\begin{table}[ht]
\centering
   \begin{tabular}{|c|c|c|c|c|}
    \hline
    $K_t$ & $K_s$ & $K_p$ & $K_h$ & $K_{sw}$ \\
    \hline
    130 \kJm & 16.6 \Nm & 3.5 \Nm & 5 \kJm & $2\cdot10^4$ \kJm\\
    \hline
    \hline
    $g_t$ & $g_s$ & $g_p$ & $g_h$ & $g_{sw}$ \\
    \hline
    0.58 & 1.62  & 1 & 0.02 & 100 \\
    \hline
   \end{tabular}
   \caption{Values of the coupling constants (above) and of their normalized counterpart (below).}
   \label{tab:dynParms}
\end{table}
%
\par
For numerical calculations it is more convenient to renormalize the Lagrangian 
with respect to some energy value in order to work with dimensionless quantities.
To be coherent with the \Y\ work we renormalized the Lagrangian with respect 
to the pairing coefficient $l_p=K_p(d_{bs}+2r)^2/2\simeq223\, \kJm$; the values of renormalized 
coefficients for coupling constants $g_t=K_t/l_p$, $g_s=K_s(d_{bs}+r)^2/2/l_p$, $g_h=K_h/l_p$
and $g_{sw}=K_{sw}/l_p$ are summarized in table~\ref{tab:dynParms}.
\par
Comparing equation~\ref{eq:gyak} with the potential used by \Y\ in~\cite{Yak04} 
it is easy to see that the coupling constant values induced on the \Y\ model are given by
\eq
g = 2g_t+4(1+\alpha)^2g_s\simeq30\;,\;\;K=2(1+\beta)^2g_p\simeq5.8
\feq
so that the normalized \Y\ coupling constant (the torsional one) corresponding to our choice 
of parameters is $g/K\simeq5.1$.
\par
Note that, when using the Morse expression for pairing, the constant in front of the pairing
is simply the bond energy $D=0.1\, \kJm$ (see eq.~(\ref{eq:Vp})). We decide nevertheless
to divide by the same normalizing factor $l_p=K_p(d_{bs}+2r)^2/2=D\mu^2$ so that all  
dimensionless coupling constants are unchanged except for the pairing one which becomes 
$g_p'=1/\mu^2\simeq0.013$.
\subsection{Dispersion relations}
The linearized version of the composite Lagrangian is
\eqn
L_{\hbox{lin}}(\theta_{n,i},\varphi_{n,i},\dot\theta_{n,i},\dot\varphi_{n,i}) =&\
&\phantom{+ }\displaystyle\sum_{n=1}^{N}\sum_{i=1}^2 \left\{\frac{1}{2}I_t \dot\theta^2_{n,i} + \frac{1}{2}I_s\left[\dot\varphi_{n,i}+(1+\alpha)\dot\theta_{n,i}\right]^2\right\}\cr
&&- \displaystyle\sum_{n=1}^{N-1}\sum_{i=1}^2\frac{1}{2} K_t \left[\Delta\theta_{n,i}\right]^2\cr
&&- \displaystyle\sum_{n=1}^{N-1}\sum_{i=1}^2\frac{1}{2} K_s(d_{bs}+r)^2\left[\Delta\varphi_{n,i}+(1+\alpha)\Delta\theta_{n,i}\right]^2\cr
&&- \displaystyle\sum_{n=1}^{N}\frac{1}{2} K_p(d_{bs}+2r)^2\left[\varphi_{n,1}+\varphi_{n,2}+(1+\beta)(\theta_{n,1}+\theta_{n,2})\right]^2\cr
&&- \displaystyle\sum_{n=1}^{N}\sum_{i=1}^2\frac{1}{2} K_h [\theta_{n+5,i+1}-\theta_{n,i}]^2
\feqn
leading to the following equations of motion:
\eq
\begin{cases}
I_t\ddot\theta_{n,i}+(1+\alpha)I_s[\ddot\varphi_{n,i}+(1+\alpha)\ddot\theta_{n,i}]=&-K_t\square\theta_{n,i}\cr
&-K_s(d_{bs}+r)^2(1+\alpha)[\square\varphi_{n,i}+(1+\alpha)\square\theta_{n,i}]\cr
&-K_p(d_{bs}+2r)^2(1+\beta)\left[\varphi_{n,1}+\varphi_{n,2}+(1+\beta)(\theta_{n,1}+\theta_{n,2})\right]\cr
&-K_h\left[-\theta_{n+5,i+1}+2\theta_{n,i}-\theta_{n-5,i+1}\right]\cr
I_s[\ddot\varphi_{n,i}+(1+\alpha)\ddot\theta_{n,i}]=&-K_s(d_{bs}+r)^2[\square\varphi_{n,i}+(1+\alpha)\square\theta_{n,i}]\cr
&-K_p(d_{bs}+2r)^2\left[\varphi_{n,1}+\varphi_{n,2}+(1+\beta)(\theta_{n,1}+\theta_{n,2})\right]\cr
\end{cases}
\feq
where $\square\theta_{n,i}=-\theta_{n+1,i}+2\theta_{n,i}-\theta_{n-1,i}$ and similarly for $\varphi$.
What we are interested in are the conditions for the existence of wave solutions for the equations above, i.e. in the form
$$
\theta_{n,j}=\Theta_{\omega,k,j}\,e^{i(k\delta n+\omega t)}\;,\;\;\varphi_{n,j}=\Phi_{\omega,k,j}\,e^{i(k\delta n+\omega t)}\;,\;\;j=1,2
$$
In the approximation $\alpha\simeq\beta$, namely considering the bases as pointlike, it is easy to 
simplify a pair of equations by multiplying the second equation by $(1+\alpha)$ and substracting it from 
the first one so that we are left with the pair
$$
I_t\ddot\theta_{n,i}=-K_t\square\theta_{n,i}-K_h\left[-\theta_{n+5,i+1}+2\theta_{n,i}-\theta_{n-5,i+1}\right]\;,\;\;i=1,2
$$
Summing and subtracting and imposing the wave form for the angles on the two equations above we find,
by imposing the condition for the existence of non-trivial waves, the first pair of dispersion relations:
\eqn
\label{eq:dr1}
\omega_1^2&=&4(K_t/I_t)\sin^2(k\delta/2)+2(K_h/I_t)[1+\cos(5k\delta)]\cr
\noalign{\vskip.2cm}
\omega_2^2&=&4(K_t/I_t)\sin^2(k\delta/2)+4(K_h/I_t)\sin^2(5k\delta/2)
\feqn
The second pair of the dispersion relations can then be easily extracted from the remaining pair of equations 
after the change of coordinates $\psi_{n,i}=\varphi_{n,i}+(1+\alpha)\theta_{n,i}$, the approximation 
$d_{bs}+2r\simeq d_{bs}+r$ (which in turn implies $\alpha=\beta$) and finally the fact that $I_s/(d_{bs}+r)^2$ 
is the mass $m_b$ of the base; the pair of equations then become
$$
\begin{cases}
m_b\ddot\psi_{n,1}=&-K_s\square\psi_{n,1}-K_p\left[\psi_{n,1}+\psi_{n,2}\right]\cr
m_b\ddot\psi_{n,2}=&-K_s\square\psi_{n,2}-K_p\left[\psi_{n,1}+\psi_{n,2}\right]\cr
\end{cases}
$$
Restricting the equations on the wave solutions and summing and subtracting them we find the other pair 
of dispersion relations
\eqn
\label{eq:dr2}
\omega_3^2&=&4(K_s/m_b)\sin^2(k\delta/2)\cr
\noalign{\vskip.2cm}
\omega_4^2&=&4(K_s/m_b)\sin^2(k\delta/2)+2K_p/m_b
\feqn
Physically, the four dispersion relations correspond to the four
oscillation modes of the system in the linear regime. 
The relations involving $\omega_1$ and $\omega_2$ are associated
with torsional oscillations of the backbone. In case of $\omega_1$
there is a threshold for the generation of the excitation
originating in the helicoidal interaction, whereas the second
torsional mode $\omega_2$ has no threshold and is thus also of
acoustical type. 
The relation involving $\omega_3$ describes relative oscillations 
of the two bases in the chain with respect to the neighboring bases. 
As $\omega_3(k)\to 0$ for $k\to 0$ there is no threshold for the
generation of these phonon mode excitations.
The dispersion relation involving $\omega_4$ describes relative 
oscillations of two bases in a pair. The threshold for the generation 
of the excitation is now determined by the pairing interaction.
\par
\begin{figure} \label{fig:dispRel}
 \includegraphics[width=6cm]{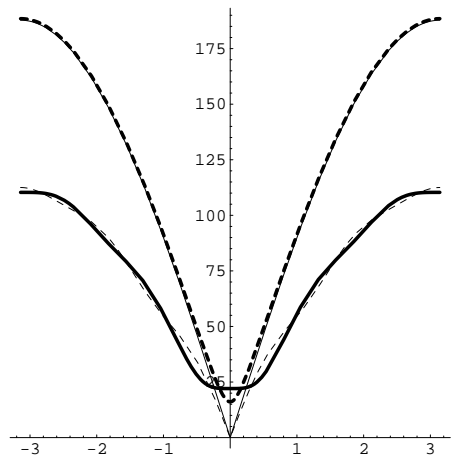}
 \caption{Graph of the dispersion relations (\ref{eq:dr1},\ref{eq:dr2})
 in the first Brillouin zone.  We plot $\omega_{\alpha}/(2\pi
 c)$ ($c$  is the speed of light) as a function of $k\delta/2$.
The $\omega_1,\omega_2,\omega_3,\omega_4$ are represented
respectively by the thick continuous, thin dashed, thin continuous and  
thick dashed line. Units are $\rm{cm^{-1}}$ in the vertical axis and radiants 
in the horizontal axis.}
\end{figure}

The dispersion relations (\ref{eq:dr1},\ref{eq:dr2}) for values of the physical
parameters given in the tables \ref{tab:kynParms} and
\ref{tab:dynParms} are plotted in Fig.~\ref{fig:dispRel}; there we
plot $\omega/(2\pi c)$, where $c$ is the speed of light (we use the,
in the literature widespread, convention of measuring frequencies
in $2\pi c$ units) versus $k\delta/2$.
\par
The four dispersion relations take a simple form if we consider
excitations with wavelength $\lambda$ much bigger then the intrapair
distance, i.e $\lambda \gg \delta$; this corresponds to the $\delta\to 0$
limit. We have then \beq \omega^2_\a \, - \, c^2_\a \, k^2 \ = \
q^2_\a \ , \feq where $c_\a$ and $q_\a$ ($\a=1\ldots 4$) are,
respectively, the velocity of propagation (in the limit $k \gg q_{\alpha}$) 
and the excitation threshold. They are given by
\beq
\lb{speed1} 
\begin{array}{ll}
  c_1 = \delta \sqrt{(K_t - 25 K_h )/I_t},\quad & q_1 = 2 \sqrt{K_h/I_t}; \\
  c_2 = \delta \sqrt{(K_t + 25 K_h )/I_t},\quad & q_2 = 0; \\
  c_3 = \delta \sqrt{K_s/m_b},\quad & q_3 = 0; \\
  c_4 = \delta \sqrt{K_s/m_b}, \quad & q_4 = \sqrt{2 K_p/m_b}; \\
\end{array}
\feq 
Using the values of the parameters given in the tables
\ref{tab:kynParms} and \ref{tab:dynParms} we have 
\beq
\label{speed}
\begin{array}{ll}
  c_1 \simeq 0 \, \kms, \quad & q_1 \simeq 4.5 \, {\rm{ cm}^{-1}} \ ; \\
  c_2 \simeq 1 \, \kms, \quad & q_2 = 0 \ ;\\
  c_3 \simeq 3 \, \kms, \quad & q_3 = 0 \ ; \\
  c_4 \simeq 3 \, \kms, \quad & q_4 \simeq 32 \, {\rm { cm}}^{-1} \ , \\
\end{array}
\feq
 where $c_1 \simeq 0$ comes from the fact that we are taking $K_t\simeq25\,K_h$
(see  table \ref{tab:dynParms} -- this of course  just means that
$c_1$ is at least an order of magnitude smaller than the other
$c_i$, and therefore negligible). Speeds can be converted to base
per seconds by dividing each $c_i$ by $\delta=3.4$\AA; excitation
thresholds can be converted in inverse of seconds by multiplying
each $q_i$ by $2\pi c$, where $c$ is the speed of light.
\par
\section{Numerical analysis}
\label{sec:numerics}
As we pointed out in sec.~\ref{sec:model}, the profiles of the solitons 
in our model are extremals of the function of $4N$ variables $L(\theta_{n,i},\varphi_{n,i})$,
where $N$ is the number of nodes of the chain. The interval of variations of
the Lagrangian parameters under our study is such that solitons can have
a radius of more than a hundred nodes so a reasonable value of $N$ is of the
order of $10^3$; following \Y~\cite{Yak04} we mostly used the value $N=2000$.
\par
Clearly the extremals -- which by the way in our case turn out to be always maxima --
of such a complex function of so many variables can only be obtained 
numerically; to accomplish that, again following \Y, we choose to use the so-called 
{\sl conjugate gradients} method.
Ready-to-use implementations of this algorithm are available in the main numerical 
libraries, in particular in Numerical Recipes (NR, \cite{NR}) and in the GNU Scientific 
Library (GSL, ~\cite{GSL}). As a double check, most of the results presented here have been 
produced using both implementations; all profiles shown in this paper 
were produced with GSL.
\subsection{Solitons profiles in the Yakushevich model}
As a generalization of the sine-Gordon solitons, also the Yakushevich solitons 
are ``relativistic'' and in particular admit a limit speed beyond which no
solution exists. We verified numerically that, for a fixed choice of the coupling
constants, solitons profiles corresponding to all possible speeds vary by less than 
$10\%$ from each other, so that it is enough to show the profiles for a single value
of the speed. All profiles presented here are relative to static solitons 
-- in other words non-trivial equilibria position of the double chain; in this case
the kynetic term is zero and the only parameter left in the renormalized Lagrangian
is the coupling constant $g$ of the torsion potential.
\par
As starting point for the extremizing algorithm, following \Y, we used
\eq
\label{eq:yakic}
\theta_{n,i}=q_i\pi\left\{1+\tanh\left[M(2n-N)\right]\right\}
\feq
where $(q_1,q_2)$ is the topological type of the soliton and $M$ a parameter 
that dictates the steepness of the profile which we use to probe different 
regions of the phase space. 
\par
\begin{center}
  \begin{figure}
    \includegraphics[width=530pt,bb = 0 310 595 775]{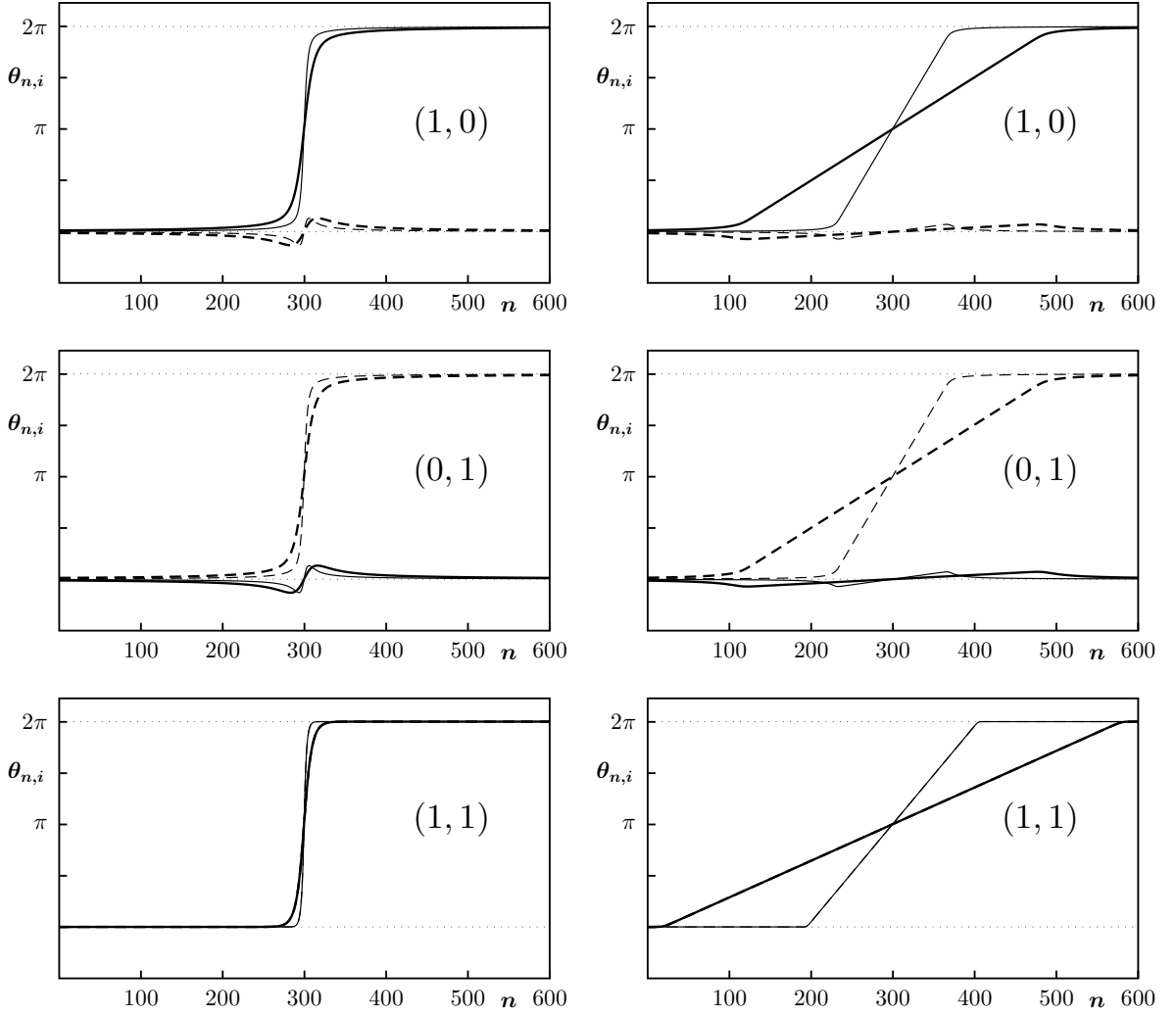}
    \caption{%
      Profiles of the solitons with the lowest non-trivial topological
      numbers for the \Y\ model in the harmonic approximation for pairing
      (left column) and with a Morse pairing potential (right). In each
      picture the thin line correspond to the physical value for the 
      coupling constant $g=21$ and the thick one, to show how the soliton 
      profile changes with $g$, to $g=150$; the continous line is relative 
      to the angles $\theta_{n,1}$, the dashed one to the angles $\theta_{n,2}$.
      In the $(1,1)$ pictures the dashed lines are not visible because the
      angles profiles for the two angles are identical.
    }
    \label{fig:yak}
  \end{figure}
\end{center}
\begin{table}
  \centering
  \begin{tabular}{|c|c|c|c|c|c|c|}
  \hline
  &\multicolumn{3}{|c|}{Harmonic}&\multicolumn{3}{|c|}{Morse}\cr
  \hline
  $(q_1,q_2)$& $g$ & $E$ (MJ/mol)& $D$ (nodes)& $g$ & $E$ (MJ/mol)& $D$ (nodes) \cr
  \hline
    $(1,0)$ & \multirow{3}{*}{21} & 4.6 & 43 & \multirow{3}{*}{29} & 0.7 & 144\cr  
    $(0,1)$ &                     & 4.6 & 43 &                     & 0.7 & 144\cr  
    $(1,1)$ &                     & 14 & 14 &                     & 1.1 & 196\cr  
  \hline
\end{tabular}
  \caption{%
    Energies and diameters for soliton profiles shown in fig.~\ref{fig:yak}; 
    diameters are defined as the number of consecutive nodes between the
    angles $\theta_{min}=1/10$ and $\theta_{max}=2\pi-\theta_{min}$.
  }
\end{table}
A first important feature of the model is the disappearance of the discrete
solitons when the coupling constant $g$ is smaller than some value $g_{min}\simeq10$
(see table~\ref{tab:instab}), way before the natural threshold represented by
when the solitons become so steep that jump from $0$ to $2\pi$ in a space
shorter than the chain step $\delta$.
This behaviour is due to the impossibility
of balancing between torsion and pairing when $g$ is too small: indeed both 
contributions ``telescope'' but the torsion ones are {\sl a priori} bounded.
A simple way to see how this happen is to give a look to the simplest case,
namely the one corresponding to topological numbers $(1,1)$, when the discrete 
Lagrange equations can be reduced to a sort of ``discrete sine-Gordon'' equations, 
namely $$g(\sin\Delta\theta_{i}-\sin\Delta\theta_{i-1})=K\square\theta_i$$
Summing term by term the first $n$ equations we get
$$g(\sin\Delta\theta_{n}-\sin\Delta\theta_{0})=K(\theta_{n+1}-\theta_{n}-\theta_{1}+\theta_{0})$$
which, considering that $\theta_0=0$ and that for $n$ of the order of $N/2=1000$
also $\theta_1\simeq0$, reduces to $\sin\Delta\theta_{n}=(K/g)\Delta\theta_{n}$.
Clearly when $g$ decreases the right term increases and when it becomes bigger
than 1 there can be no solution anymore.
\par
\begin{table}
  \centering
  \begin{tabular}{|c|c|c|c|}
  \hline
    & (1,0) & (0,1) & (1,1) \\
  \hline
  $g_0$ & 7.05 & 7.05 & 14.7 \\
  \hline
\end{tabular}
  \caption{The transition values $g_0$ for solitons instability 
    (solitons arise only for $g > g_0$) of the $(p,q)$ solitons.}\label{tab:instab}
\end{table}
This feature is particularly relevant for our study because the value $g=5.4$
induced by the values of the coupling constants we extracted from literature
(see previous section) unfortunately falls in the range where no soliton arise 
-- at least for the basic topological numbers $(1,0)$, $(0,1)$ and $(1,1)$ -- 
in the harmonic approximation for pairing. Since nevertheless $5.4$ is rather 
close to the boundary values for the existence of solutions and the estimate 
of coupling constants is only qualitative, for the harmonic approximation we
increased the value by a factor 4. Note that though this behaviour disappears
when a more physical potential, i.e. a Morse one, is used for pairing,
leading to the conclusion that the feature above belongs to the harmonic
version of the model but does not play any role in realistic models of DNA.
\par
In fig.~\ref{fig:yak} we show the results for the solitons corresponding to 
the basic topological numbers $(1,0)$, $(0,1)$ and $(1,1)$ both in the harmonic 
approximation for pairing (used in \Y\  in~\cite{Yak04}) and with a Morse potential 
for two values of the coupling constant $g$, one corresponding to the physical
value and one corresponding to a value about an order of magnitude above in order
to enhance the detail in the solitons profile (whose diameter increases monotonically 
with $g$).
As expected, the profiles corresponding to the Morse potential have a bigger radius,
actually by about an order of magnitude, than the corresponding harmonic ones.
In all cases the numerical solutions appear to be very robust with respect to 
the initial configuration, so that we can affirm that in all of them there is
a unique extremal.
\subsection{Solitons profiles in the composite model}
\begin{figure}
  \includegraphics[width=200pt,bb = 170 530 470 800 ]{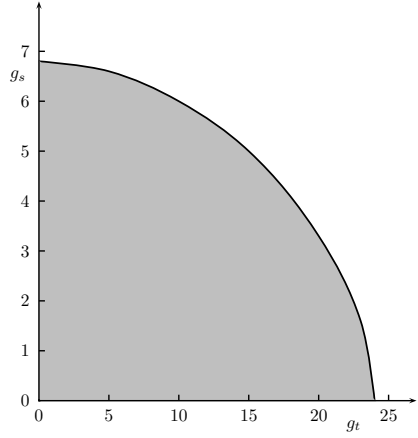}
  \caption{%
    Region of the $(g_t,g_s)$ plane where no solitons arise for the $(0,1)$ topological type.
  }
  \label{fig:gtgs}
\end{figure}
\begin{center}
  \begin{figure}
    \includegraphics[width=530pt,viewport = 0 310 595 775,clip]{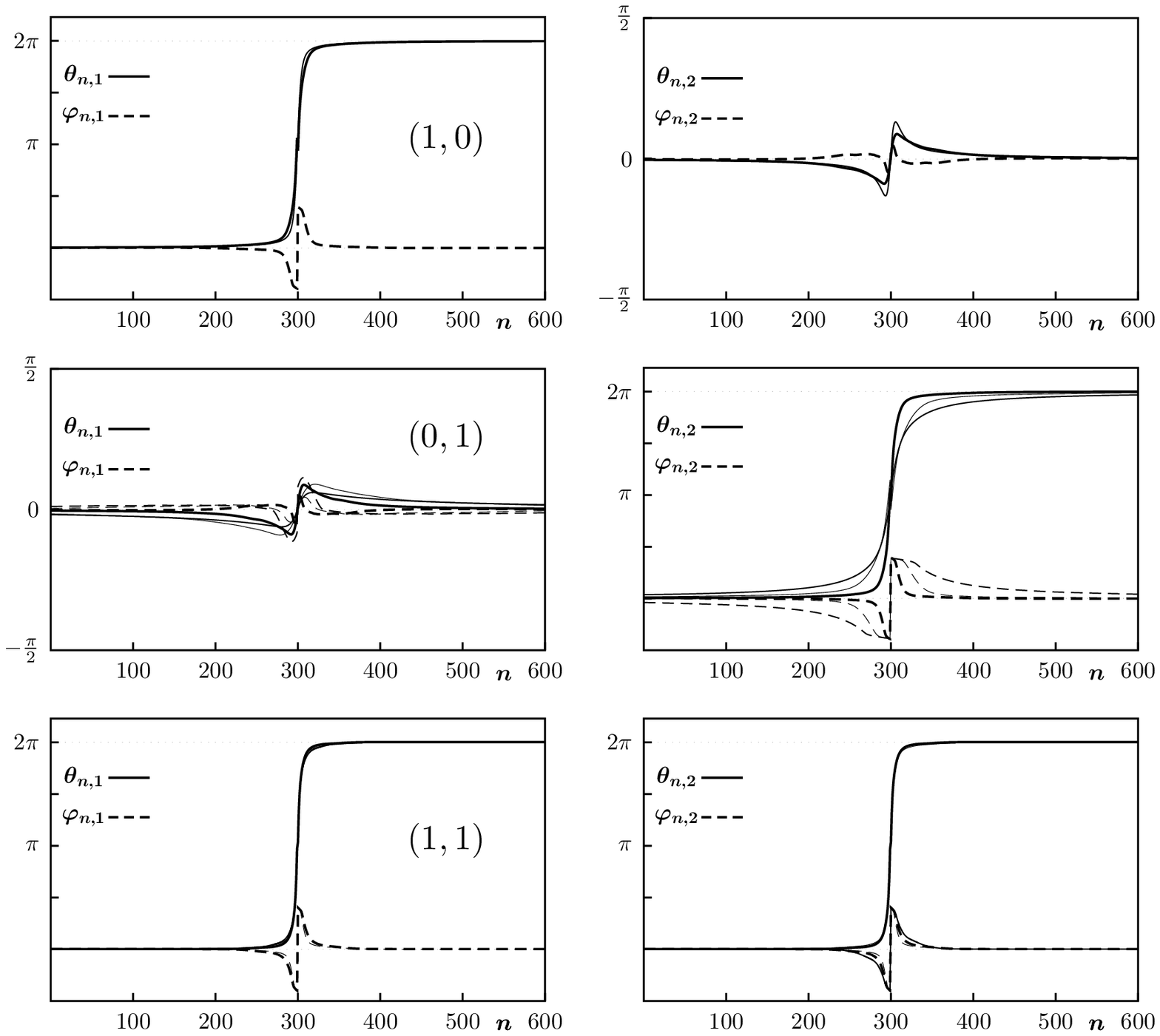}
    \caption{%
      Profiles of solitons corresponding to the smallest non-trivial topological numbers
      for the composite model in the harmonic approximation for pairing.
      In the $(1,0)$ case we compare the profile with the corresponding \Y\ one to 
      evidence how close the topological component in the composite model are to
      their \Y\ counterpart.
      In the $(0,1)$ case we compare the profiles to the ones we get by increasing the
      torsional part to values corresponding to $g=150$ in \Y\ model, once by
      putting the whole contribute in $V_t$ (thin line) and then putting it wholly in $V_s$
      (thinner line). In both cases the profiles diameters get bigger, as expected, 
      but no new qualitative feature appears.
      In the $(1,1)$ case we compare the profiles to the ones we get by not making the
      contact approximation (i.e. keeping $\bho_{eq}\neq0$, thinner line) or by enlarging 
      the helicoidal coupling constant (thin line). Even in this case no new feature appears.
    }
  \label{fig:gharm}
  \end{figure}
\end{center}
\begin{table}
  \centering
  \begin{tabular}{|c|c|c|c|c|c|c|}
  \hline
  &\multicolumn{3}{|c|}{Harmonic}&\multicolumn{3}{|c|}{Morse}\cr
  \hline
  $(q_1,q_2)$& $g_s$ & $E$ (MJ/mol)& $D$ (nodes)& $g_s$ & $E$ (MJ/mol)& $D$ (nodes) \cr
  \hline
    $(1,0)$ & \multirow{3}{*}{6.5} & 21 & 54 & \multirow{3}{*}{1.6} & 0.7 & 136\cr  
    $(0,1)$ &                      & 21 & 54 &                      & 0.7 & 136\cr  
    $(1,1)$ &                      & 63 & 30 &                      & 1.1 & 154\cr  
  \hline
\end{tabular}
  \caption{%
    Energies and diameters for soliton profiles shown in fig.~\ref{fig:gharm} and~\ref{fig:gmorse}; 
    diameters are defined as the number of consecutive nodes between the
    angles $\theta_{min}=1/10$ and $\theta_{max}=2\pi-\theta_{min}$.
  }
\end{table}
As a generalization of  the sine-Gordon model, also in this composite model 
there is very little difference
between profiles corresponding to different speeds
\footnote{Notice that for the composite model the boost symmetry is realized in a highly 
non trivial way owing to the presence of two (instead of a single one) limiting speeds for the 
propagation of travelling waves (see~\cite{CDG3} for details)}; 
in particular, we will consider even in this case only stationary profiles.
\par
As starting point for the extremizing algorithm we used the natural generalization
of~(\ref{eq:yakic}):
$$
\theta_{n,i}=q_i\pi\left\{1+\tanh\left[M(2n-N)\right]\right\}\,,\;\phi_{n,i}=0\,.
$$
Since the angles $\phi_{n,i}$ are bound to only half of a circle, the corresponding
field in the continous approximation cannot describe a topologically non-trivial
path and therefore we call these coordinates ``non-topological'' and there is no
integer number associated to them, while the topological numbers $q_i$ of the $\theta$
angles correspond exactly to the numbers in the simpler model.
\par
\begin{center}
  \begin{figure}
    \includegraphics[width=530pt,bb = 0 310 595 775]{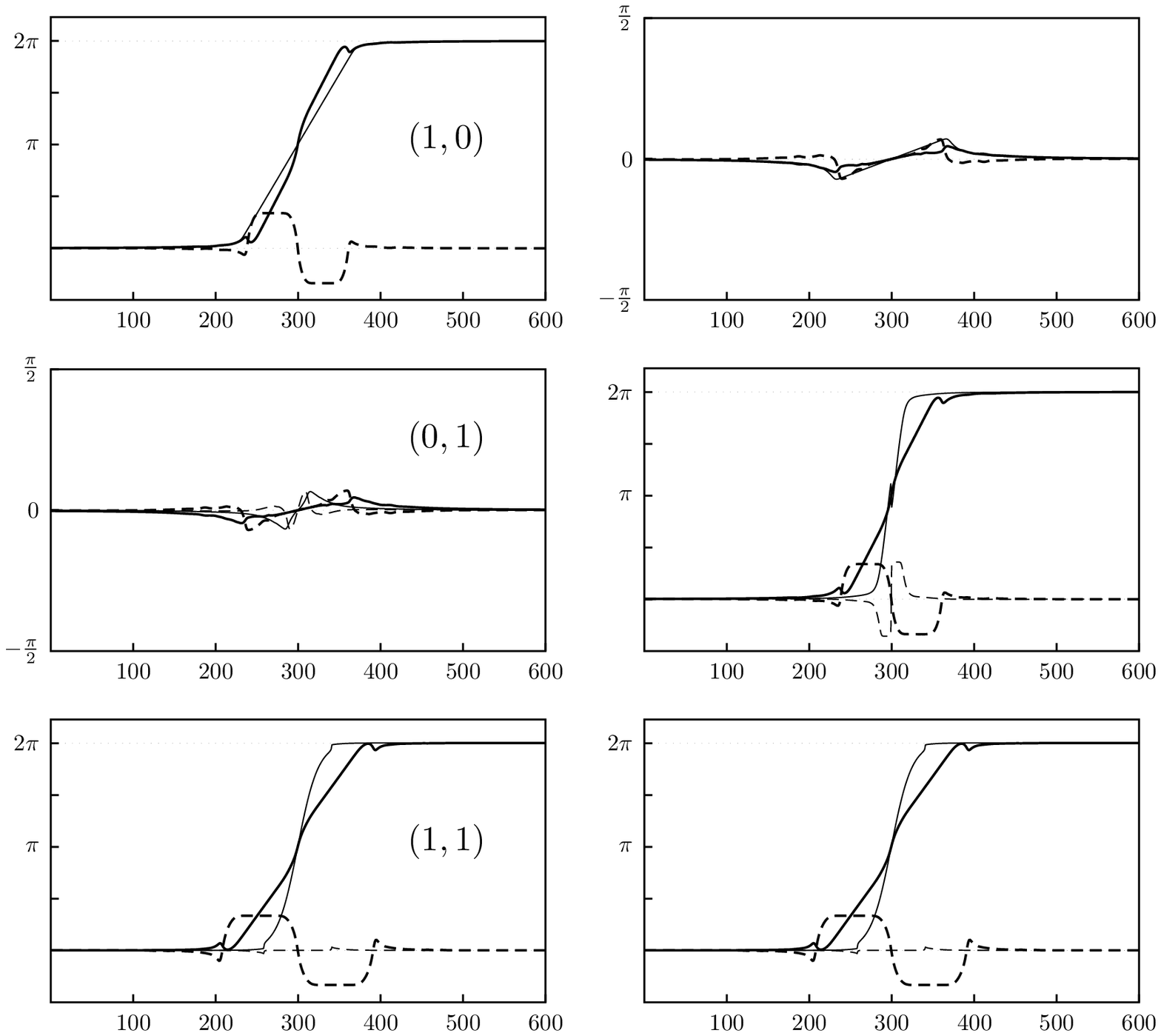}
    \caption{%
      Profiles of solitons corresponding to the smallest non-trivial topological numbers
      for the composite model with a Morse potential for pairing.
      In the $(1,0)$ case we compare the profile with the corresponding \Y\ one to 
      evidence how close are even in this case the topolgical component in the composite 
      model to their \Y\ counterpart.
      In the $(0,1)$ case we compare the profiles to the ones we get by decreasing the
      well width normalized parameter $\mu(d_{bs}+2r)$ from 8.8 to 2 (thinner line). 
      As expected, the profile with a smaller $\mu$ gets narrower, due to the fact 
      that the radius of the Morse hole got bigger and therefore the pairing interaction 
      looks more like a globally harmonic one.
      In the $(1,1)$ case we compare the profiles to the ones we get by disregarding
      completely both torsional couplings. Differently from what happens in the 
      harmonic approximation, when solitons become trivial for small torsions,
      the profile has the same qualitative features of the one corresponding to the
      physical coupling constants.
    }
    \label{fig:gmorse}
  \end{figure}
\end{center}

A further similarity between the two systems is the disappearance of the discrete soliton 
for too little values of the coupling constants, in this case $g_t$ and $g_s$ 
(see fig.~\ref{fig:gtgs} -- the helicoidal term is at least an order of magnitude smaller than them and it can
be disregarded). This is why in all profiles relative to the composite model in the
harmonic approximation (fig.~\ref{fig:gharm}) we magnified $g_s$ by a factor 4, bringing it to
$g_s=6.5$; when a Morse-like potential is used instead (fig.~\ref{fig:gmorse}) solitons 
profiles survive even when both torsion and stacking are turned off, so in that case 
we keep using the correct value $g_s=1.8$.
\par
In this section we show the profiles of the solitons for the base topological types
$(1,0)$, $(0,1)$ and $(1,1)$ for both the harmonic pairing approximation (fig.~\ref{fig:gharm}) 
and the more physical Morse potential (fig.~\ref{fig:gmorse}) comparing them with some of the 
several deformations that can be tested to check their stability.
\subsubsection{Harmonic approximation}
In fig.~\ref{fig:gharm}, first we compare the plot
of the $(1,0)$ soliton with the corresponding soliton in the \Y\ model; the two graphs 
are within $10\%$ from each other, testifying that the increase in complexity of the
model gives us more features without modifying qualitatively the successful results of the 
model that generalizes.
\par
Then we compare the profiles of the $(0,1)$ soliton with those corresponding to a torsional 
coupling an order of magnitude
bigger. Since in this model there are two distinct sources for torsion, we choose to show
the profiles corresponding to the extremal cases, namely when all torsion is concentrated in
the backbone and when it is instead concentrated in the bases; in this case there are bigger
differences between the profiles but we can say that, even after magnifying by an order of
magnitude the torsion with respect to the pairing, the picture stays qualitatively the same.
\par
Finally we compare the profiles of the $(1,1)$ soliton with those
obtained by not doing the contact approximation (i.e. by considering the equilibrium position
of bases being at distance of 3\AA\ from each other) and those obtained by magnifying the 
helicoidal contribution by a two orders of magnitude. Even in this case the differences in
the profiles turned out negligible.
\par
Summarizing the results, the harmonic approximation turned out to be rather solid even
in the composite model but it keeps suffering of the same problem already detected in 
the \Y\ model, namely there is no non-trivial discrete soliton if the torsional part
is ``too small'' and unfortunately the physical values of the coupling constants seem
to follow inside this non-existence zone, or at the very least very close to its boundary.
\par
A second concern is that the profiles of the non-topological angles, even the ones generated 
with higher torsional couplings, in correspondance to the soliton rise from 0 to $2\pi$ 
jump from about $-\pi/2$ to $\pi/2$ within a single node. This on one side casts some doubt
on the existence of a continous conterpart, leading to possible problems for their movement
evolution on the chain, and on the other side testifies of a quite violent struggle
about the points $\pm\pi/2$, where two very strong non-linear forces (the pairing and
the sugar wall) compete with each other; this behaviour is quite unwelcome first because
it is unphysical, since after the ionic hydrogen bonds break (and this happens when they 
get a few \AA\ apart) they do not interact anymore, and second because such
a violent non-linear interaction most likely destabilizes the system.
\par
\subsubsection{Morse potential}
%
In fig.~\ref{fig:gmorse}, first we compare the plot of the $(1,0)$ soliton with the corresponding soliton in the \Y\ model; 
even with the Morse potential the similarity of the profiles obtained for the two models is striking,
keeping within $10\%$ from each other.
\par
Then we compare the plot of the $(0,1)$ soliton with the ones obtained by decreasing the well
width parameter $\mu$, i.e. enlarging the well from $1$\AA\ to $4$\AA. As expected
we get a norrower profile since in the limit $\mu\to0$ the Morse potential reduces to its
harmonic approximation.
\par
Finally we show the $(1,1)$ soliton profiles and compare them with the ones obtained by
neglecting altogether both torsional couplings. The profile is of course narrower but it
is still non-trivial, as opposite to the case of the harmonic approximation when even at the
physical values for the coupling constants the only profiles we obtain are the constant
ones.
\par
Summarizing the numerical analysis relative to the Morse potential, we get the same nice
features found in the harmonic approximation but we loose its worst defects. It seems rather 
clear therefore that a serious investigation of DNA's rotational dynamical properties cannot 
avoid using a Morse-like potential for pairing, i.e. the pairing coupling must be turned 
completely off after the distance between bases increases by a few \AA.
\section{Conclusions}
The numerical analysis we have performed shows  the
existence of solitonic solutions of our composite DNA model.
The profiles of the topological solitons -- in particular, the
part relating to the topological degree of freedom -- of our model
are both qualitatively and quantitatively very similar to those of
the \Y\ model. This means that the most relevant (for DNA
transcription) and characterizing feature of the nonlinear DNA
dynamics present in the \Y\ model is preserved by considering
geometrically more complex and hence more realistic DNA models.
\par
Moreover, the topological soliton profiles of our model seem to
change very little when either the physical parameters change in a
reasonable range or also the form of the potential modelling the
pairing interaction is modified to a more realistic form with a
sole exception, namely the replacement of the pairing Morse potential 
with its harmonic approximation, which works fine close to the 
equilibrium position but fails badly when the bases flip.
\par
In particular, the forms of the topological solitons are very
little sensitive to the interchange of torsional and stacking
coupling constant.
This feature adds other reasons why the \Y\ model, although based on
a strong simplification of the DNA geometry, works quite well in
describing solitonic excitations. The \Y\ model, indeed, does not
distinguish between torsional and stacking interaction; but, as we
have shown, this distinction is not relevant -- at least as long
as one is only interested in the existence and form of the soliton
solutions.
The ``compositeness'' of our model becomes relevant -- and
rather crucial -- when it comes on the one hand to allowing the
existence of solitons {\it together} with requiring a physically
realistic choice of the physical parameters characterizing the
DNA, and on the other hand to have also predictions fitting
experimental observations for what concerns quantities related to
small amplitude dynamics, such as transverse phonons speed. In
other words, the somewhat more detailed description of DNA
dynamics provided by our model allows it to be effective -- with
the same parameters -- across regimes, and provide meaningful
quantities in both the linear and the fully nonlinear regime.
\par
We expect that the model considered here is the simplest DNA model
describing rotational degrees of freedom which, with physically
realistic values of the coupling constants and other parameters,
allows for the existence of topological solitons and at the same
time is also compatible with observed values of bound energies and
phonon speeds in DNA. We also expect that solitons may move
for considerable distances in this model even in presence of 
realistic inhomogeneities thanks to the fact that the component
that supports the solitons motion, i.e. the sugar-phosphate group,
is homogeneous and therefore by separating it from the bases we
can consider the inhomogeneities as a perturbative effect.
We leave to a future paper the study of inhomogeneities and
time evolution in this model.
\section*{Acknowledgements}
We gladly thank Giuseppe Gaeta for introducing the problem and
both Giuseppe Gaeta and Mariano Cadoni for several enlightening 
discussions on the subject and for readproofing this manuscript.
\bibliography{cshNum}
\end{document}